\documentstyle[prb,aps,twocolumn,floats,epsfig,twoside]{revtex}

\newcommand{\bea}{\begin{eqnarray}}
\newcommand{\eea}{\end{eqnarray}}
\newcommand{\be}{\begin{equation}}
\newcommand{\ee}{\end{equation}}

\begin{document}

\twocolumn[\hsize\textwidth\columnwidth\hsize\csname@twocolumnfalse\endcsname

\title{Non-equilibrium dynamics of simple spherical spin models}
\author{W. Zippold, R. K\"uhn, H. Horner\\Institut f\"ur Theoretische 
Physik,
Universit\"at Heidelberg, Philosophenweg 19, D-69120 Heidelberg}
\date{\today} 
\maketitle

\begin{abstract}
We investigate the non-equilibrium dynamics of spherical spin models with
two-spin interactions. For the exactly solvable models of the 
d-dimensional
spherical ferromagnet and the spherical Sherrington-Kirkpatrick (SK) 
model 
the asymptotic dynamics has for large times and large waiting times the 
same formal structure. In the limit of large waiting times we find in 
both models an intermediate time scale, scaling as a power of the waiting 
time with an exponent smaller than one, and thus separating the 
time-translation-invariant   short-time dynamics from the aging  regime. 
It is this
time scale on which the fluctuation-dissipation theorem is violated.  
Aging in these
models is similar to that observed in spin glasses at the level of 
correlation
functions, but different at the level of response functions, and thus 
different at
the level of experimentally accessible quantities like  thermoremanent
magnetization.

\end{abstract}
\pacs{PACS numbers: 75.10.Hk, 75.10.Nr, 05.50.+q, 64.60.Fr}

]

\section{Introduction}

There exist many systems which exhibit relaxation times long enough to
keep them from reaching equilibrium on experimental time scales. Primary
examples are spin glasses and polymer glasses, but also systems as simple
as the Ising model when prepared in an arbitrary initial state, or phase 
separation dynamics in systems with conserved parameter such as Ostwald 
ripening in binary alloys. As a consequence, the relaxation depends 
for all these system on the waiting time $ t_w $ already spent in the low 
temperature phase: the systems age. To understand the aging phenomena 
observed in these models one has to investigate their non-equilibrium 
dynamics.

In this context the investigation of the non-equilibrium dynamics of 
spherical spin models with two-spin interactions is interesting because 
they
exhibit nontrivial dynamical behaviour, despite their simplicity, which 
makes their
non-equilibrium dynamics exactly  solvable. These systems never  reach 
equilibrium,
hence correlation and response functions depend on the waiting  time even 
in the
limit of large times \cite{curev,bray,cude}. 

Our main aim is to complete for this class of models the analysis of 
the spherical SK model presented in \cite{cude} by identifying 
{\em all\/} relevant time scales of the problem. We are going to show 
that, in 
addition to the two time regimes found in \cite{cude}, there exists an 
intermediate time scale $ t_p \gg 1 $ satisfying $ t_p/t_w \to 0 $ for
$t_w\to\infty$. It is this intermediate time scale on which the 
fluctuation
dissipation theorem is beginning to be violated. Interest in this time 
scale stems
from the  fact that a thorough understanding of the dynamics at these 
intermediate
times  is important with respect to the study of the non-equilibrium 
dynamics 
in models more complicated than those considered in the present paper, 
such 
as that of the spherical p-spin glass with $ p > 2 $, because it is 
the 
behaviour at the time scale $ t_p $ which determines the behaviour at the 
time scale $ t_w $ in a unique way. It is thus the key ingredient towards 
the solution of the so far unsolved problem of selecting a unique 
solution 
within an infinite family of time reparametrization covariant solutions 
on 
diverging time scales, as has been demonstrated within a multi-domain 
crossover scaling approach for the closely related problem of a slowly 
dragged 
particle in a random potential \cite{hordr}. The analysis of the simple 
spherical spin models is presented here, because their behaviour at the 
intermediate time scale can be studied analytically and in instructive 
detail.

Moreover we shall see that, despite the similarity of these models to
the more difficult case of the spherical p-spin glass, their dynamics is 
not
spin glass dynamics. This has been realized for some time from 
considerations
concerning fluctuation dissipation ratios or parametric plots of an 
integrated 
response versus correlation (see e.g. \cite{curev,cukupe}). 
Alternatively, 
one may look at the thermoremanent magnetization (another form of 
integrated 
response) as a quantity sensitive to the complicated phase space 
structure, 
to distinguish spin glasses from the simpler magnetic systems. While the 
thermoremanent magnetization, when plotted against logarithmic time, 
exhibits 
a waiting time dependent plateau in spin glasses, this plateau is absent 
in 
the models considered here. 

We have organized our material as follows. In Sec. II we introduce the 
models
and briefly review the general method for solving their non-equilibrium 
dynamics,
as first presented in \cite{cude}. In Sec. III we specialize to the 
spherical
SK model and to $d$ dimensional hyper-cubic spherical ferromagnets, which 
independently of the dimension $d$ of the latter, exhibit formally the 
same 
type of long time non-equilibrium dynamics; exponents describing the 
decay of 
correlation and response for the latter vary, of course, with $d$. Time 
scales
are identified and analyzed in Sec. IV, while Sec. 5 contains a 
discussion of
our results. 

\section{The model}
We consider spherical spin models with two spin interactions consisting of
$ N $ continuous spins $ s_i(t) $, $ i= 1,\ldots,N $, which satisfy for 
all 
times $ t $ the spherical constraint $ \sum_{i=1}^N {s_i}^2(t) = N $. The 
Hamiltonian of the system is given by
\be\label{spham}
H = -\frac{1}{2}\sum_{i\neq j}J_{ij}s_i s_j\ .
\ee

The coupling matrix $ J_{ij} $ is supposed to be an arbitrary symmetric
matrix. Denoting the eigenvalues of the matrix $ J $ by $ a_i $, 
$ i = 1,\ldots,N $, the system of Langevin equations, which describes the 
dynamics of the model, decouples in terms of the projections $ s_{a_i}(t) 
$ 
of the spins $ s_i(t) $ onto the eigenvectors 
\bea
\label{langevinsp}
\partial_t s_{a_i}(t) = (a_i - \mu(t))s_{a_i}(t) + h_{a_i}(t) &&  + 
\xi_{a_i}(t)\ ,
\nonumber\\ && i = 1,\ldots,N\ , 
\eea
where $ h_{a_i}(t) $ is the corresponding component of an external 
magnetic
field and $ \xi_{a_i}(t) $ is thermal Gaussian white noise with zero mean
and correlation $ \langle\xi_{a_i} (\tau + t_w)\xi_{a_j} (t_w)\rangle 
= 2T\delta_{ij}\delta (\tau) $. The parameter $ \mu(t) $ is the Lagrange 
multiplier enforcing the spherical constraint. Henceforth we will use 
$ \langle\cdot\rangle $ to represent the average over the thermal noise. 
If 
it were not for the Lagrange parameter $ \mu(t) $, the dynamics 
(\ref{langevinsp}) would just be that of $ N $ independent harmonic 
oscillators under the influence of thermal noise. This means that solving 
the 
non-equilibrium dynamics of these models reduces to determining $ \mu(t) 
$. 
It was shown in reference \cite{cude} that for a given waiting time $ t_w 
$ 
and given time separation $ \tau\geq 0 $ the autocorrelation 
$ q(\tau,t_w) := 1/N\left[\sum_{i=1}^N\langle s_i(\tau + t_w)s_i(t_w)
\rangle\right]_J $ and response function $ r(\tau,t_w) := 1/N
\left[\sum_{i=1}^N  \delta \left.\langle s_i(\tau+t_w)\rangle/
\delta h_i(t_w)\right|_{h=0}\right]_J $ of this class of models are in 
terms
of
\be\label{deflambda}
\Lambda(t) := \exp\left( 2\int_0^t ds\,\mu(s)\right) 
\ee
given by 
\bea\label{qsp}
q(\tau,t_w) =&& 
\frac{\Lambda(t_w+\tau/2)}{\sqrt{\Lambda(\tau+t_w)\Lambda(t_w)}}\ \Bigg[1 
           - \\
&&T\int_0^\tau ds\,\frac{\Lambda(t_w+\tau/2-s/2)}{\Lambda(t_w+\tau/2)}\,
            \langle\langle\exp(as)\rangle\rangle\Bigg]\nonumber
\eea
and
\be\label{rsp}
r(\tau,t_w) = \sqrt{\frac{\Lambda(t_w)}{\Lambda(\tau+t_w)}}\,
\langle\langle\exp(a\tau)\rangle\rangle\ ,
\ee
where we have specialized the expressions in \cite{cude} to the case of 
zero
external field and constant temperature $ T $ and have chosen the initial 
condition to be $ s_{a_i}(t=0) = 1 $. By $ \left[\cdot\right]_J $ we have
denoted a possible disorder average and by 
$ \langle\langle\cdot\rangle\rangle $
we denote the integration $ \int da\,\rho(a)\cdot $ over an eigenvalue 
density 
$ \rho(a) $ which in the thermodynamic limit $ N\to\infty $ describes the 
distribution of eigenvalues of the coupling matrix $ J $. The 
quantity $ \Lambda(t) $ itself is determined by 
\be 
\label{lambda}
\Lambda(t) = \langle\langle \exp(2at)\rangle\rangle + 2T\int_0^t ds\, 
\Lambda(s)\langle\langle\exp(2a(t-s))\rangle\rangle\ ,
\ee
which together with (\ref{qsp}) immediately implies 
$ \Lambda(0) = q(0,t_w) = 1 $. Another dynamical observable we will be 
interested in is the thermoremanent magnetization $ m_r(\tau,t_w) $. 
Given that 
the system is kept in a small magnetic field $ h $ in the time interval 
$ [0,t_w] $ , the magnetization measured at time $ \tau + t_w $ is given 
by
\be\label{defrm}
m_r(\tau,t_w) = h\int_0^{t_w}ds\,r(\tau+s,t_w-s)\ .
\ee 
For models such as the $d$ dimensional ferromagnets considered in what 
follows, 
in which the interaction matrix has a geometrical structure, off-diagonal
correlations of the form $q_{ij}(\tau , t_w) = \langle s_i(\tau+t_w) 
s_j(t_w) 
\rangle$ are of course also of interest. We have not investigated them in 
the present paper, however, as our main interest here is in results which 
will have further bearing on spin glass models of the mean field type.

\section{Spherical SK model and spherical ferro\-magnet}

While the expressions given so far are valid for any choice of the 
coupling
matrix $ J $ we now want to treat two special cases. Our aim is to solve
the non-equilibrium dynamics of these particular models in the limit of 
large 
waiting times $ t_w \gg1 $ by explicitly determining $ \Lambda(t) $. As 
we are
only interested in the behaviour of the dynamical observables for times
$ \tau,t_w \gg 1 $ it is sufficient to determine the asymptotic behaviour 
of
$ \Lambda(t) $ for $ t \gg 1 $. In the following we will discuss the 
d-dimensional spherical ferromagnet and the spherical SK model. The 
latter 
is the special case $ p=2 $ of the disordered spherical p-spin model and
we will present the results found in \cite{cude} in a slightly different 
form.

In the case of the spherical ferromagnet we consider a $ d $-dimensional
hyper-cubic lattice with periodic boundary conditions, whose lattice
constant we take to be unity and whose lattice sites with coordinate 
vectors
$ \vec{x}_i $ are occupied by spins $ s_i $. The couplings are chosen to 
be
ferromagnetic nearest neighbour interactions, whose strength is set to 
unity. 
The standard diagonalization procedure using Fourier modes \cite{bax} 
yields for this choice of the matrix $ J $ in the limit $ N\to\infty $ 
for 
the spectrum of eigenvalues and the eigenvalue density $ \rho^{fm}(a) $ 
the 
result 
\bea\label{dichtefm}
\rho^{fm}(a) = \frac{1}{\pi}\int_0^{\infty}dy\, \cos(ay) && 
\left[J_0(2y)\right]^d\ ,\nonumber\\
&&\ a\in [-2d,2d]\ ,
\eea 
where $ J_0(y) $ denotes the Bessel function of zeroth order.

The spherical SK model is defined by choosing the coupling matrix $ J $ to
be a random matrix whose entries $ J_{ij} $ are independent and 
identically
distributed Gaussian random variables with zero mean and variance
$ [(J_{ij})^2]_J = 1/N $. A general result of random matrix theory 
\cite{meht} states that, for
this choice of $ J $, the eigenvalue density in the thermodynamic limit 
$ \rho^{sk} $ is given by the Wigner semi-circle law
\be\label{dichtesk}
\rho^{sk}(a) = \frac{1}{2\pi}\sqrt{4 - a^2}\ ,\quad a\in[-2,2]\ .
\ee

Solving the non-equilibrium dynamics of these models means solving 
(\ref{lambda}) for the eigenvalue densities (\ref{dichtefm}) and 
(\ref{dichtesk}). This is best done using the Laplace transform
$ \tilde{\Lambda}(s) = \int_0^{\infty}dt\,\Lambda(t)\exp(-st) $
to obtain from (\ref{lambda}) for $ t > 0 $ the relation
\be\label{laplam}
\tilde{\Lambda}(s) = \frac{\tilde{f}(s)}{1-2T\tilde{f}(s)}\ ,
\ee
where the function $ \tilde{f}(s) := \left\langle\left\langle 
1/(s-2a)\right\rangle\right\rangle $ is characteristic of the given 
model. In terms of the 
function $ f(t) $, which yields $ \tilde{f}(s) $ via Laplace 
transformation, 
the expressions (\ref{qsp}) and (\ref{rsp}) can be rewritten as
\bea\label{qfmsk}
q(\tau,t_w) =&& 
\frac{\Lambda(t_w+\tau/2)}{\sqrt{\Lambda(\tau+t_w)\Lambda(t_w)}}\ \Bigg[1 
          \\ - 
&&\frac{1}{2}T\int_0^{2\tau} dx\,
\frac{\Lambda(t_w+\tau/2-x/4)}{\Lambda(t_w+\tau/2)}\,f(x/4)\Bigg]\ .
\nonumber
\eea
and
\be\label{rfmsk}
r(\tau,t_w) = \sqrt{\frac{\Lambda(t_w)}{\Lambda(\tau+t_w)}}\,f(\tau/2)\ ,
\ee
respectively. Inserting the expressions for the eigenvalue densities
(\ref{dichtefm}) and (\ref{dichtesk}) in the definition of $ \tilde{f}(s) 
$
we find that in the case of the spherical ferromagnet the function $ f(t) 
$ 
is given by
\be\label{ftfm}
f^{fm}(t) = [I_0(4t)]^d\ ,
\ee
while for the spherical SK model it is calculated to be
\be\label{ftsk}
f^{sk}(t) = \frac{I_1(4t)}{2t}\ .
\ee
In these expressions $ I_0(t) $ and $ I_1(t) $ denote the modified Bessel
function of zeroth and first order, respectively. 

The critical temperature of the dynamic phase transition is found from 
(\ref{laplam}) to be 
\be\label{tcsp}
T_c = \frac{1}{2\tilde{f}(2a_m)}\ ,
\ee
where $ a_m $ denotes the maximal eigenvalue of the eigenvalue spectrum 
$ [-a_m,a_m] $ of the corresponding model. In the special cases 
considered 
here we have $ a_m^{fm} = 2d $ and $ a_m^{sk} = 2 $. Expression 
(\ref{tcsp}) implies in the case of the spherical ferromagnet that we get 
a 
phase transition in $ d > 2 $ only. For the spherical SK model the 
critical
temperature can be calculated explicitly and one finds $ T_c^{sk} = 1 $. 
These results are in both models in agreement with the ones obtained from 
static calculations of the transition temperature \cite{bk},\cite{ktj}.

In the following we will only be interested in the low temperature phase 
with
$ T < T_c $. In this phase the asymptotic behaviour of $ \Lambda(t) $ 
for large times $ t $ is determined by the behaviour of the Laplace 
transform
$ \tilde{\Lambda}(s) $ at the right bound $ s = 2a_m $ of the branch cut.
In the case of the spherical SK-model the inverse Laplace transformation 
can 
be done exactly and the result found in \cite{cude} reads
\be\label{lambdask}
\Lambda^{sk}(t) = \frac{1}{T}\sum_{k=1}^{\infty} kT^k\,\frac{I_k(4t)}{2t}\ .
\ee
For the spherical ferromagnet there is no simple analytic expression, but 
we 
can find the leading asymptotic behaviour for $ t\gg 1 $ in integer 
dimension 
$ d > 2 $ by expanding $ \tilde{\Lambda}(s) $ around $ s = 2a_m $ and 
performing the inverse Laplace transformation. Performing this 
calculation 
and comparing the result with the leading order of expression 
(\ref{lambdask}) 
for large times $ t\gg 1 $, we find that in both models the leading 
asymptotic 
behaviour of $ \Lambda(t) $ for $ t\gg 1 $ can be written as
\be\label{asylambda}
\Lambda(t) \simeq \frac{\Lambda_0}{(1-T/T_c)^2}\frac{e^{2a_mt}}{t^{\nu_s}}
\hspace{1cm}t\gg 1\ ,
\ee
where the prefactor $\Lambda_0$ is given by $ \Lambda_0^{fm} = 
(8\pi)^{-\nu_s} $ for the spherical ferromagnet and by $ \Lambda_0^{sk} = 
(32\pi)^{1-\nu_s} $ for the spherical SK model. The exponent $ \nu_s $ 
appearing in these expressions is $ \nu_s^{fm} = d/2 $ for the spherical 
ferromagnet and $ \nu_s^{sk} = 3/2 $ for the spherical SK model. In the 
same way it follows from (\ref{ftfm}) and (\ref{ftsk}) that the 
asymptotic 
behaviour of $ f(t) $ can in both models be written as
\be\label{asyft}
f(t) \simeq \Lambda_0 \frac{e^{2a_mt}}{t^{\nu_s}}\hspace{1cm}t\gg 1\ ,
\ee
with the same factor $ \Lambda_0 $ as in (\ref{asylambda}). These two 
formulas 
indicate already the close correspondence of the asymptotic behaviour of 
autocorrelation and response function in both models which we will study 
in more detail in the following section. We will see that we can write 
the 
expressions for these dynamical observables for both models in a unified 
way 
using the exponent $ \nu_s $ defined above. This means in particular that 
we 
will find the same time scales appearing in the limit of large waiting 
times 
$ t_w $ and large time separations $ \tau $.

\section{Asymptotic dynamics and time scales}

Before entering the discussion of the asymptotic behaviour for large
waiting times $ t_w $, we want to present for later reference the 
expressions 
for correlation and response function for finite waiting time $ t_w \sim 
1 $ 
and large times $ \tau \gg 1 $. Using (\ref{asylambda}) and (\ref{asyft}) 
in the
terms containing the time variable $ t $ in (\ref{qfmsk}) and 
(\ref{rfmsk}) 
we obtain for the autocorrelation
\be\label{kurztwq}
q(\tau,t_w) \sim f_q(t_w)\tau^{-\nu_s/2}\hspace{1cm}\tau\gg t_w\simeq 1\ ,
\ee
where $ f_q(t_w) \simeq 1 $ for $ t_w \sim 1 $, and for the 
response
\be\label{kurztwr}
r(\tau,t_w) \simeq f_r(t_w)\tau^{-\nu_s/2}\hspace{1cm} \tau \gg t_w\simeq 
1\ ,
\ee
where again $ f_r(t_w) \sim 1 $.

We are, however, mainly interested in the case of large waiting times 
$ t_w \gg 1 $ and large time separations $ \tau \gg 1 $. In this case 
we can insert the asymptotic expansions (\ref{asylambda}) and 
(\ref{asyft})
in expressions (\ref{qfmsk}) and (\ref{rfmsk}) and get for $ \tau,t_w \gg 
1 $ 
\bea\label{asyq}
q(\tau,t_w) && \simeq 
\frac{\left(1+\frac{\tau}{t_w}\right)^{\nu_s/2}}{\left(1
+\frac{\tau}{2t_w}\right)^{\nu_s}}
\times \\ && 
\Bigg[1-\frac{1}{2}T\int_0^{2\tau}dx\,e^{-a_mx/2}\frac{f(x/4)}{\left(1
-\frac{x}{4t_w(1+\tau/2t_w)}\right)^{\nu_s}}\Bigg]  \nonumber
\eea
for the leading asymptotic behaviour of the autocorrelation and 
\be\label{asyr}
r(\tau,t_w) \simeq 
b\left(1+\frac{\tau}{t_w}\right)^{\nu_s/2}\,\tau^{-\nu_s}
\ee
for the response.
The prefactor $ b $ is $ b^{fm} = (4\pi)^{-\nu_s} $ in the case of the
ferromagnet and $ b^{sk} = (4\pi)^{1-\nu_s} $ for the SK model. These
equations show that the asymptotic dynamics in the limit $ \tau,t_w \gg 1 
$
of the spherical ferromagnet and the spherical SK model possesses the same
{\em formal\/} structure. In particular we find that the scaling
behaviour of the dynamical observables autocorrelation and response of 
the 
spherical ferromagnet in $ d=3 $ and the spherical SK model is equivalent
(on the level of exponents). This result was independently stated in 
\cite{curev}. The formal correspondence of the two models implies in 
particular that in both models the same characteristic time scales appear.

Before entering the discussion of the relevant time scales in the problem,
let us simplify expression (\ref{asyq}) further. Expanding the 
denominator of
the integrand in this expression in a power series, one can prove that in
the limit of large waiting times $ t_w \gg 1 $ the dominant contribution 
to
the integral comes for all times $ \tau \gg 1 $ from the zeroth order 
term of
the expansion. Defining
\be\label{qeasp}
q_p := 1-\frac{T}{2}\int_0^{\infty}dx\,e^{-a_mx/2}f(x/4) = 1 - 
\frac{T}{T_c}\ ,
\ee
where the last equality follows from (\ref{tcsp}), we find
\be\label{leadasyq}
q(\tau,t_w) \simeq 
\frac{\left(1+\frac{\tau}{t_w}\right)^{\nu_s/2}}{\left(1
+\frac{\tau}{2t_w}\right)^{\nu_s}}\left(q_p + c_0\tau^{1-\nu_s}\right)
\ee
for the leading behaviour of the autocorrelation in the limit $ \tau,t_w 
\gg 1 $, 
in which $c_0 = b T/(\nu_s -1)$.

Using (\ref{asyr}) and (\ref{leadasyq}) it is now straightforward
to identify the different time scales of the problem. At first sight
we find the two time scales already discussed
in \cite{cude} for the case of the spherical SK model. The first is the
time scale $ t_0 \sim 1 $ of the microscopic relaxation. At the upper end
of this scale we have $ \tau \gg 1 $ but still $\tau/t_w \ll 1 $, such 
that we
can neglect all waiting time dependent corrections. On this time scale the
dynamics corresponds to the dynamics in equilibrium, i. e. it is time 
translation invariant with autocorrelation $ q(\tau,t_w) = 
\tilde{q}_0(\tau_0)$
and response $ r(\tau,t_w) = \tilde{r}_0(\tau_0) $ being functions of
the scaling variable $ \tau_0 := \tau/t_0 $ only, and autocorrelation and 
response satisfy the FDT $ -\partial_{\tau} \tilde{q}_0(\tau_0) 
= T\tilde{r}_0 (\tau_0) $ of equilibrium dynamics. Therefore we will 
refer to 
this time scale as the FDT regime. At the upper end $ \tau_0 \gg 1 $ of 
this 
scale the response is found from (\ref{asyr}) to be
\be\label{fdtr}
r(\tau,t_w)  = \tilde{r}_0(\tau_0) \simeq \hat b_0\,\tau_0^{-\nu_s}\ ,
\ee
with $\hat b_0 = b t_0^{-\nu_s}$, while (\ref{leadasyq}) implies for the 
correlation
\be\label{fdtqsp}
q(\tau) = \tilde{q}_0(\tau_0) \simeq q_p + \hat c_0\tau_0^{1-\nu_s}
\ee
with $\hat c_0 = c_0 t_0^{1-\nu_s}$. This corresponds to a power law 
decay of 
the correlation to a plateau value $ q_p $, which in the case of the 
spherical 
ferromagnet is just the square of the static spontaneous magnetization 
$\langle s_i\rangle^2$ \cite{bk}, while in the case of the spherical SK 
model 
it is the static Edwards-Anderson parameter $ q_{EA} = \left[\langle s_i 
\rangle^2\right]_J $ \cite{cude,ktj}. The exponent of the decay is
\be
\nu_0 := 1-\nu_s
\ee
which in the case of the spherical SK model is just the special case $ 
p=2 $
of the result found in \cite{hcsed} for the equilibrium decay of the 
correlation of the spherical p-spin glass. If we speak of a plateau in the
correlation, it is of course understood that this plateau in the 
autocorrelation is only visible in a plot against the logarithm of time 
$\tau$. 

The second obvious time scale is the waiting time itself. For $ \tau \sim 
t_w $ 
correlation and response can be written as functions of the scaling 
variable 
$ \tau_w := \tau/t_w $ and
one finds asymptotically for $ \tau_w \gg 1 $ power law decays of the
dynamical observables to zero \cite{cude}.

From (\ref{asyr}) it is obvious that these two are the only time scales
which can be identified from the behaviour of the response function. 
However,
it turns out that there exists a further nontrivial time scale in the
problem, which can be identified from the autocorrelation function. 
Looking at
expression (\ref{leadasyq}) and taking into account the leading waiting 
time 
dependent correction in the prefactor we arrive at
\be\label{compasyq}
q(\tau,t_w) \simeq 
\left(1-\frac{\nu_s}{8}\left(\frac{\tau}{t_w}\right)^2\right)
\left(q_p + c_0\tau^{1-\nu_s}\right)\ .
\ee
This expression shows that there exists a waiting time dependent scale 
$ t_p(t_w) $, on which the correlation begins to decay away from the 
plateau 
value $ q_p $. To be more precise, we define this time scale $ t_p $ by
requiring $ q(t_p,t_w) = q_p $, such that $ t_p $ corresponds to the 
middle
of the plateau of the correlation function. This means that $ t_p $ is 
the 
time for which the competing corrections in (\ref{compasyq}) are of the 
same 
order of magnitude. Hence we find that the time scale $ t_p(t_w) $ scales 
as
\be\label{tp}
t_p(t_w) \sim {t_w}^{2/(1+\nu_s)}\ll t_w
\ee
with the waiting time $ t_w $. The latter inequality follows as $ \nu_s > 
1 $.
The plateau regime corresponding to time scale $ t_p $ is the so far 
missing 
link between the stationary dynamics within the FDT regime and the 
non-stationary dynamics for times of the order of the waiting time $ t_w 
$ 
itself. We will shortly see that it is in particular the time scale on 
which 
the FDT of equilibrium dynamics is violated.

In terms of the scaling variable $ \tau_p := \tau/t_p $, the correlation 
within 
the plateau regime $ \tau \sim t_p $ can be expressed in the scaling form 
\be\label{qanstp}
q(\tau,t_w) = q_p + \hat{q}_p(t_w)\tilde{q}_p(\tau_p)
\ee
with the prefactor $\hat{q}_p(t_w) = t_w^{-2(\nu_s - 1)/(\nu_s + 1)} \sim 
t_p^{\nu_0}$ and the scaling function
\be
\tilde{q}_p(\tau_p) \simeq  c_0 \tau_p^{\nu_0}  + c_p{\tau_p}^{\nu_1}
\label{obtpq} 
\ee
in which $ c_p = -\frac{\nu_s}{8}\,q_p $ and $ \nu_1 = 2 $ (recall that
$\nu_0 = 1-\nu_s < 0$). This scaling function describes the decay of the 
correlation  towards $q_p$ ad the lower end of the plateau scale, i.e. 
for 
$\tau_p \ll 1$, and its subsequent decay away from $ q_p $ at the upper 
end
of the plateau scale where $\tau_p \gg 1$.

In order to study the violation of the FDT we introduce a quantity 
$n(\tau,t_w)$ which characterizes this violation quantitatively \cite{hordr} 
via
\be\label{defn}
-\partial_\tau q(\tau,t_w) =: T\,(1+n(\tau,t_w))r(\tau,t_w)\ .
\ee
Note that differentiating with respect to the time separation is 
equivalent
to differentiating with respect to the later time $t$. Alternatively one 
may
differentiate with respect to the earlier time $t_w$. In the 
representation
of the correlation in terms of $t_w$ and time difference $\tau = t - t_w$ 
this
gives rise to the corresponding definition
\be\label{defnw}
\hat\partial_{t_w} q(\tau,t_w) =: T\,(1+n_w(\tau,t_w))r(\tau,t_w)\ ,
\ee
with $\hat\partial_{t_w} = \partial_{t_w} -\partial_\tau$. The latter is
related to the fluctuation dissipation ratios $X(\tau,t_w)$ studied e.g. 
in 
\cite{curev,cukupe} via $X(\tau,t_w) = 1/(1+n_w(\tau,t_w))$. For 
$\tau,t_w 
\gg 1$ we get
\bea
1+n(\tau,t_w)  && \simeq \frac{1}{\left(1 
+\frac{\tau}{2t_w}\right)^{\nu_s}}
\times \label{noftau}
\\ && \Bigg[1 + \frac{\nu_s q_p}{4bT}\frac{\tau^{1+\nu_s}}{t_w^2} \frac{1}
{\left(1 +\frac{\tau}{t_w}\right)\left(1 +\frac{\tau}{2t_w}\right)} \Bigg]\ ,
\nonumber\\
1+n_w(\tau,t_w) &&  \simeq \frac{1}{\left(1 
+\frac{\tau}{2t_w}\right)^{\nu_s}}
\times \label{nwoftau} \\
&& \Bigg[1+ \frac{\nu_s q_p}{4bT}\frac{\tau^{1+\nu_s}}{t_w^2} \frac{1}
{\left(1 +\frac{\tau}{2t_w}\right)} \Bigg]\ . \nonumber
\eea
Fig. 1 shows these two functions for $T=0.6\,T_c$, and $t_w = 10^{10}$ so that $t_p=10^8$.


\begin{figure}
[ht]
{\centering 
\epsfig{file=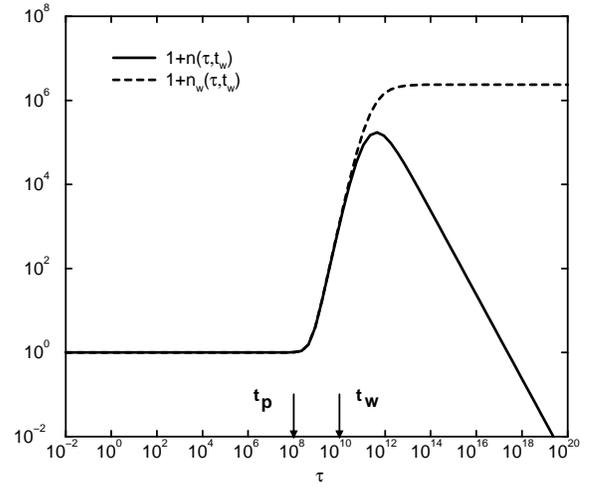,width=0.500\textwidth}  
\par}
\caption{$1+n(\tau,t_w)$ and \  $1+n_w(\tau,t_w)$ \ as functions of $\tau$ for $t_w=10^{10}$. Vertical arrows mark the plateau scale $t_p(t_w)$ and the
waiting time scale $t_w$.}
\vspace{0.2truecm}
\label{figure}
\end{figure}


As we have seen before, the decay towards the plateau satisfies the FDT, 
that is, we have in leading order $ n(\tau,t_w) = n_w(\tau_tw) = 0 $  for 
$ \tau \ll t_p $. On the intermediate time scale $ t_p = 
t_w^{2/(1+\nu_s)}$, 
however, we obtain the scaling form 
\be
n(\tau,t_w) \simeq n_w(\tau,t_w) \simeq\tilde{n}(\tau_p) =
\frac{\nu_s q_p}{4 b T}\,{\tau_p}^{1+\nu_s}\ ,
\ee
which approaches zero at the lower end of the plateau scale (where the FDT
holds) but is non-zero (indicating FDT violation) for all $\tau_p = {\cal 
O}(1)$, 
and exhibits a power law divergence at the upper end of the $t_p$ 
scale. This can be traced back to the fact that the behaviour of the 
response 
function $ r(\tau,t_w) $ does not change on the time scale $ t_p $ 
whereas 
that of the correlation does. It is this divergence which is responsible 
for 
the fact noted in \cite{curev,cukupe} that parametric representations of 
integrated response $\chi(\tau,t_w) = \int_0^\tau d s r(s,t_w +\tau -s)$ 
versus correlation $q(\tau,t_w)$ saturate at the value $\chi = (1- 
q_p)/T$ for 
$q(\tau,t_w) \le q_p$ for the models considered in the present paper. 
Note 
that the nature of this divergence is in the large $t_w$ limit of course 
not 
detectable in such parametric plots, as it occurs entirely on the plateau
scale, on which $q(\tau,t_w)$ is basically arrested at $q_p$. It may 
however 
be obtained from the finite $t_w$-corrections to such plots which may be 
extracted from (\ref{noftau}) and (\ref{nwoftau}).

Note that (\ref{noftau}) and (\ref{nwoftau}) imply that $n$ and $n_w$ 
exhibit
{\em different\/} scaling on the $t_w$ scale. With $\tau_w = \tau/t_w$ we 
have
\bea
1+n(\tau,t_w) && \simeq \frac{1}{\left(1 +\tau_w/2\right)^{\nu_s}}
\times \label{nnoftau}\\
&& \Bigg[1+ \frac{\nu_s q_p}{4bT}\,t_w^{\nu_s-1} \tau_w^{1+\nu_s} \frac{1}
{\left(1 +\tau_w\right)\left(1 +\tau_w/2\right)} \Bigg]\ ,\nonumber\\
1+n_w(\tau,t_w)  && \simeq \frac{1}{\left(1 +\tau_w/2\right)^{\nu_s}}
\times \label{nnwoftau}\\
&& \Bigg[1+ \frac{\nu_s q_p}{4bT}\,t_w^{\nu_s-1} \tau_w^{1+\nu_s} \frac{1}
{\left(1 +\tau_w/2\right)} \Bigg]\ ,\nonumber
\eea
implying that both are infinite in the $t_w\to \infty$ limit, but show 
different behaviour at large $\tau_w$ when $t_w$ is large but finite: 
Whereas 
$1 + n(\tau,t_w) \simeq \frac{\nu_s q_p}{bT}\, (2 t_w)^{\nu_s-1}\, 
\tau_w^{-1} 
\to 0$ as $\tau_w \to \infty$, we have $1+ n_w(\tau,t_w) \simeq  
\frac{\nu_s 
q_p}{bT}\, (2 t_w)^{\nu_s-1}$ in the same limit.

To summarize, the FDT is broken already on the time scale $ t_p $ rather 
than only on the scale $ t_w $, the former being much smaller than the 
latter when $ t_w $ becomes large, since $ t_p/t_w \to 0 $ as $ t_w \to 
\infty $. Moreover, the divergence of $n$ and $n_w$ on the $t_p$ scale 
implies that the QFDT solution, which was found in \cite{hcsed} for the 
spherical p-spin glass with $ p >2 $ and in \cite {hk} for manifolds in
disordered potentials, does not exist in the case of the spherical SK 
model
(and in the ferromagnetic systems). From (\ref{compasyq}) it is also 
obvious that for the models considered here the plateau regime $ t_p $ is 
the only further time scale in the problem. This, too, is in contrast to 
the expectations for spherical p-spin glass with $ p >2 $, which will 
be considered in a forthcoming paper \cite{neqps}. 

\section{Discussion}

Considering the simplicity of the models discussed in the previous 
sections,
the complexity of the dynamical behaviour seems rather astonishing. 
However, in the case of spherical ferromagnet the explicit dependence on 
the
waiting time of both correlation and response even in the limit 
$ \tau\gg t_w\gg 1 $ is a well 
known result in the theory of phase ordering kinetics \cite{bray}. 
According 
to the scaling hypothesis of coarsening dynamics there exists for large 
times 
$ t =\tau + t_w \gg 1 $ a single length scale $ L(t) $ in the system, 
which can 
be interpreted as the typical size of a domain at time $ t $. This means 
that 
for large times $ t\gg t_w \gg 1 $ the two-time-autocorrelation function 
of 
such a system is a function of the ratio of the two length scales 
$ L(t)\gg L(t_w)\gg 1 $ only. The exact solution of the phase ordering 
dynamics of the spherical ferromagnet yields for the autocorrelation at 
$ T=0 $ in the limit of large times the result \cite{bray}
\be\label{phoqasy}
q(t-t_w,t_w) = \left(\frac{4t t_w}{(t +t_w)^2}\right)^{d/4}\ ,
\ee
which is just equation (\ref{asyq}) for $ T=0 $.

It has been noted that aging behaviour in the correlation functions of
coarsening systems has a simple interpretation in terms of domain growth
\cite{bray,curev,cukupe,be+}. This holds in particular for the emergence 
of 
the plateau scale. After the system has spent the waiting time $ t_w\gg1$ 
in
the low temperature phase, an arbitrary spin will on average be found in a
domain of size $ L(t_w) $. The autocorrelation of this spin will for short
times decay towards $ q_p $, which is the square  of the local 
spontaneous 
magnetization, due to spin fluctuations within this domain. This decay is 
equivalent to a decay within a local equilibrium state and satisfies time 
translational invariance and FDT. The further asymptotic decay of the 
autocorrelation away from this value $ q_p $ towards zero can only be 
produced by a change of the environment of the chosen spin, which means 
that a domain wall has to pass by its site. As the size of the original 
domain grows as a power of the waiting time, it is very plausible that 
the 
time spent near the plateau value should also grow as a power of the 
waiting time. 
Since the growth of the domains and therefore the wandering of the 
domain walls is a slow process the asymptotic decay towards zero is also 
a 
slow power law decay. For the spherical SK model, such arguments are of 
course not available, as the model does not possess a geometry.

At the heart of it, the {\em formal\/} equivalence of the asymptotic 
dynamics 
for $ \tau,t_w \gg 1 $ of the spherical ferromagnet and the spherical SK 
model 
is due to the fact that both interaction matrices exhibit eigenvalue 
densities 
$\rho(a)$ whose behaviour at the upper end $a_m$ of the spectrum can be 
characterized by a power law
\be
     \rho(a) \sim (a_m - a)^{\nu_s - 1}\ ,\quad {\rm as} \quad a\to a_m\ ,
\ee
with the exponent $\nu_s$ introduced earlier. It is this feature which
determines the behaviour of correlation and response in these systems at
$\tau,t_w \gg 1$. The origin of the power law may be disorder, as in the 
case
of the SK model and the semi-circle law, but it need not, as exemplified 
by the 
$d$-dimensional ordered systems. Thus aging in the spherical SK model 
cannot 
be interpreted as spin glass aging as it is observed experimentally 
\cite{sveno,aohrm,vinsu} as well as in model calculations \cite{cuku1}
and simulations \cite{rie}. Indeed, it is well known that from a static 
point 
of view this model does not have the properties of a typical spin glass, 
as 
it has a replica symmetric solution for all temperatures and does not 
possess 
many degenerate ground states. The results above imply that the spherical 
SK 
model is  neither a spin glass from a dynamical point of view, despite 
the 
existence of a plateau in the correlation function as it is observed in 
spin 
glasses and related systems \cite{hcsed,hordr}. 

Obviously the autocorrelation is not a suitable quantity to distinguish 
aging
in a spin glass from the simpler case of coarsening dynamics in magnetic 
systems, whose nonequlibrium dynamics is determined by domain growth. A 
dynamical observable which characterizes a spin glass, however, is given 
by 
the thermoremanent magnetization defined in (\ref{defrm}). This is due to 
the 
fact that it is the response function which is most sensitive to the 
complex 
phase space structure exhibited by spin glasses. To be more precise, the
particular metastable configurations of a spinglass depend strongly on a 
magnetic
field. During the waiting time the system is expected to move to 
configurations of
increasing stability. On the other hand, a state which is relatively 
stable in a
given field might become less stable if the field is slightly changed. 
This means
that, after a change of the field at $t_w$, the system has to move to new 
states
of increasing stability. The time scale of this process depends on the 
degree of
stability reached at $t_w$. This leads to a plateau in the thermoremanent
magnetization similar to the one found in the correlation function. This 
will
be derived for the spherical $ p $-spin glass with $ p >2 $ in a 
forthcoming paper
\cite{neqps}.  A mechanism of this kind is of course absent in a 
coarsening system
and as a consequence  $ m_r(\tau,t_w) $ decays in the limit of large 
waiting times $
t_w \gg 1 $ for all $ \tau\ll t_w $ as
\be\label{shortrm}
m_r(\tau,t_w) \sim \tau^{1-\nu_s}. 
\ee
To prove this result, let us denote by $ t_1(t_w) $ a lower bound of the 
waiting time scale satisfying $ \tau \ll t_1\ll t_w $. Let us 
further choose a time $ t_2 $, such that $ t_w -t_2 \sim 1 $. Then we can 
split the integration in (\ref{defrm}) as follows
\bea\label{rmsplit}
m_r(\tau,t_w) \simeq && h\Big(\int_0^{t_1}ds\,r(\tau + s,t_w)\\
&& + \int_{t_1}^{t_2}ds\,r(s,t_w-s)
+\int_{t_2}^{t_w}ds\,r(s,t_w-s)\Big).\nonumber
\eea
Using (\ref{fdtr}) in the first integral we find that this term yields the
leading order contribution given in (\ref{shortrm}) as the contribution 
from 
the upper bound is negligible in the limit $ t_w \gg 1 $. In the last 
integral
in (\ref{rmsplit}) the argument $ t_w-s $ is always of order unity and 
with 
(\ref{kurztwr}) we find that it scales as $ {t_w}^{-\nu_s/2} $ with the
waiting time, such that it is negligible in the limit of large waiting 
times.
Thus we just have to consider the contributions from the middle of the 
integration range for the remaining integral in (\ref{rmsplit}). Rewriting
this integral in terms of the scaling variable $ \sigma := s/t_w $ 
we get using (\ref{asyr}) that this term scales as $ {t_w}^{1-\nu_s} $ 
which
leads to (\ref{shortrm}) as the dominant contribution. Hence we have 
indeed
found that in the type of models considered here the thermoremanent
magnetization does not exhibit a plateau in the limit $ t_w\gg 1 $ nor 
does it depend on the waiting time for all times $ t\ll t_w $.  For 
coarsening systems this is what we expected as this relaxation stems 
from spin fluctuations within a certain
domain, which do not know anything about the waiting time. As noted in 
\cite{curev,cukupe} an alternative criterion to distinguish aging in 
coarsening 
systems from spin glass aging is the integrated response $\chi(\tau,t_w)$ 
 mentioned 
in Sec. IV. Both, the saturation of $\chi(\tau,t_w)$ and the absence
of a plateau in the thermoremanent magnetization are due to the same 
reason,
namely due to the divergence of the function $n(\tau,t_w)$, equivalently 
due
to the vanishing of the fluctuation dissipation ratio $X(\tau,t_w)$ on 
the plateau scale.

Let us finally stress that the time scale $ t_p $  also appears in the 
more
complicated case of the spherical $ p $-spin glass with $ p >2 $, the
spherical SK model being just the simplest of this class of models, and 
it is
the behaviour of correlation and response on this time-scale which is 
needed
to uniquely fix the dynamics at later times. This will be explicitly 
shown in 
a forthcoming paper \cite{neqps}. 

\acknowledgements
It is a pleasure to thank H. Kinzelbach for numerous illuminating discussions.

\vfill
\eject

\end{document}